\documentclass[onecolumn,showpacs,aps,amssymb,floatfix,prd,amsmath,preprintnumbers]{revtex4}
\usepackage{epstopdf}
\usepackage{capt-of}
\usepackage{graphicx}  
\usepackage{dcolumn}   
\RequirePackage{graphicx}
\RequirePackage{float}
\RequirePackage{hyperref}
\RequirePackage{amsmath}
\RequirePackage{amssymb}
\RequirePackage{mathtools}  

\begin{document}

\title{\bf Accelerating bianchi type dark energy cosmological model with cosmic string in $f(T)$ Gravity}
\author{ S. H. Shekh$^{1}$\footnote{Email:  da\_salim@rediff.com},
  V. R. Chirde $^{2}$\footnote{Email: vrchirde333@rediffmail.com},}\
  
\affiliation{$^{1}$ Department of Mathematics, S.P.M. Science and Gilani Arts and Commerce College, Ghatanji, Yavatmal, Maharashtra-445301, India.}
\affiliation{$^{2}$ Department of Mathematics, G.S.G. Mahavidyalaya, Umarkhed-445206, India.}

\begin{abstract} 
\textbf{Abstract:} In this paper, we have investigated some features of anisotropic accelerating Bianchi type-I cosmological model in the presence of two non-interacting fluids, i.e. one usual string and other dark energy fluid towards the gravitational field equations for the linear form of $f(T)$ gravity, where \textit{T  }be the torsion. To achieve a physically realistic solution of the field equations, we have considered an exact matter dominated volumetric power law expansion. The aspects of derived model are discussed with the help of solution and make the model at late times turn out to be flat Universe. Also, observed that the model has initially non-singular and stable while for whole expansion it is unstable. At an initial expansion $0.01\le t\le 0.05$ when the Universe start to expand to infinite expansion $t>1.73$ (break away from for small interval of cosmic time $0.06\le t\le 1.73$), there is a matter dominated era while for $0.06\le t\le 1.73$ it approaches the quintessence region. Also, the string exists in the early universe which is in agreement with constraints of CMBR data but with the expansion the string phase of the Universe disappears, i.e. we have an anisotropic fluid of particles.
\end{abstract}


\maketitle

\section{Introduction}
The Cosmic Microwave Background radiation as seen in data from satellites such as the Wilkinson Microwave Anisotropy Probe (Knop \textit{et al.} 2003) and from observations of Large Scale Structure  (Tegmark \textit{et al.} 2004a, 2004b) of the universe confirmed that the problem of cosmic acceleration one. Basically, two kinds of alternative explanations have been proposed for this unexpected observational phenomenon. One is, the problem lies in detecting an exotic type of unknown repulsive force (an exotic energy with large negative pressure), termed as dark energy which is responsible for the accelerating phase of the Universe and other, by modifying the Einstein equation called Modified Gravity which is of great importance because it can successfully explain the rotation curve of galaxies and the motion of galaxy clusters in the universe. \\
The detection of dark energy would be a new clue to an old puzzle: the gravitational effect of the zero-point energies of particles and fields (Zel'dovich 1968). The total with other energies, that are close to homogeneous and nearly independent of time, acts as dark energy. The paramount characteristic of the dark energy is a constant or slightly changing energy density as the Universe expands, but we do not know the nature of dark energy very well (Turner \& Huterer 2007; Bartelmann 2010). Dark energy has been conventionally characterized by the Equation of State parameter $\omega =p/\rho $, where \textit{$\omega$ } is not necessarily constant. The \textit{$\omega$ }lies close to $\mathrm{-}$1: it would be equal to (standard cosmology), a little bit upper than $\mathrm{-}$1 (the quintessence dark energy) or less than $\mathrm{-}$1 (phantom dark energy). The possibility $\omega <<-1$ is ruled out by current cosmological data. Akarsu \& Kilinc (2010) have studied locally rotationally symmetric Bianchi type-I cosmological models with a constant deceleration parameter in the presence of anisotropic dark energy and a perfect fluid. They have considered phenomenological parameterization of minimally interacting dark energy in terms of its equation of state parameter and time-dependent skewness parameters. Pawar \& Solanke (2014a) gives an exact kantowski-sachs anisotropic dark energy cosmological models in Brans Dicke theory of gravitation. Pawar \textit{et al.} (2014b) investigated magnetized dark energy cosmological models with time dependent cosmological term in Lyra geometry. Mahanta et al. (2014) constructed dark energy anisotropic Bianchi type-III cosmological models with a variable equation of state parameter~in Barber's second self-creation theory of gravitation using the special law of variation of Hubble's parameter that yields a constant value of the deceleration parameter and obtained two different models in which first  model the value of~equation of state parameter~is in good agreement with the recent observations of type Ia supernovae data with cosmic microwave background radiation anisotropy and galaxy clustering statistics. Recently Aditya et al. (2019) investigated the dark energy phenomenon by studying the Tsallis holographic dark energy within the framework of Brans Dicke scalar--tensor theory of gravity where the Brans and Dicke scalar field~is a logarithmic function of the average scale factor and Hubble horizon as the IR cutoff and very recently Singh et al. (2020) examined dark energy cosmological model in modified scale covariant theory of gravitation. The dynamical dark energy models are classified into two different categories: (i) the scalar fields including Quintessence, Phantom, Quintom, K-essence, Tachyon, Dilaton, and so forth; (ii) the interacting models of dark energy such as the Chaplygin gas models (Sarkar  2014), Agegraphic (Wei \& Cai 2008) and Holographic (Thomas 2002; Setare 2007) models. In recent years, an interesting observation is made to determine the nature of dark energy in quantum gravity which is termed as holographic dark energy using principle of holographic dark energy. This principle (the degrees of freedom in a bounded system should be finite and does not scale by it volume but with its boundary era) was first put forwarded by Hooft (1993) in the context of black hole physics. Latter on Fischler \& Susskind (1998) have proposed a new version of the holographic principle, viz. at any time during cosmological evolution, the gravitational entropy within a closed surface should not be always larger than the particle entropy that passes through the past light-cone of that surface. In the context of the dark energy problem, though the holographic principle proposes a relation between the holographic dark energy density $\rho _{\Lambda } $ and the Hubble parameter $H$as $\rho _{\Lambda } =H^{2} $ , it does not contribute to the present accelerated expansion of the universe along with Granda \& Olivers (2008) proposed a holographic density of the form $\rho _{\Lambda } \cong \alpha H^{2} +\beta H$ , where \textit{$\alpha$, $\beta$} are constants which must satisfy the restrictions imposed by the current observational data. \\
There are various modified gravity, One replaces the Ricci scalar R  in the Einstein-Hilbert action by an arbitrary function of R  belongs to the well-known  $f(R)$ modified gravity (Azadi et al. 2008; Sharif \& Yousaf 2014; Chirde \& Shekh 2016a; 2017). Another one that has gained much regard in the last few years is Gauss-Bonnet gravity say $f(G)$ theory of gravity (Nojiri \& Odintsov 2005, 2007; Cognola 2007;  Fayaz et al. 2015; Abbas et al. 2015; Sharif \& Fatima 2016) where  $f(G)$ is a generic function of the Gauss-Bonnet invariant \textit{G}. Also the theory which combines Ricci scalar and Gauss-Bonnet scalar called $f(R,G)$ gravity theory (Alvaro \& Diego 2012; Shekh \& Chirde 2019; Makarenko et al. 2013; Atazadeh \& Darabi 2014; Laurentis \& Revelles 2015; Shekh et al. 2020). One the gravitational action includes an arbitrary function of the Ricci scalar and trace of the stress energy tensor known as $f(R,T)$ gravity. Several authors such as Sahoo \& Mishra (2014), Chirde \& Shekh (2015, 2016b), Pawar \& Agrawal (2017) who have investigated the aspects of cosmological models in this gravity.  Among the various modifications of Einstein's theory, another one way to look at the theory beyond the Einstein equation is the Teleparallel Gravity which uses the Weitzenbock connection in place of the Levi--Civita connection and so it has no curvature but has torsion which is responsible for the acceleration of the Universe. Some relevant works in this gravity are the graphical representation of k-essence with the help of equation of state parameter describe by Sharif \& Rani (2011). In $f(T)$ gravity the existence of relativistic stars investigated by Bohmer \textit{et al.} (2011). Chirde \& Shekh (2014) have discussed some cosmological model with different sources. Gamal \& Nashed (2015) investigated anisotropic models with two fluids in linear and quadratic forms of $f(T)$gravitational models. Recently, Bhoyar \textit{et al.} (2017) discussed stability of accelerating Universe with linear equation of state parameter in $f(T)$ gravity using hybrid expansion law. \\

Bianchi type cosmological models are significant in the sense that these are homogeneous and anisotropic. Bianchi type space times provide spatially homogeneous and anisotropic models of the universe as compared to the homogeneous and isotropic FRW models, from which the course of action of isotropization of the universe is well thought-out the passage of time. The simplicity of the field equations and relative ease of the solutions of Bianchi space times are useful in constructing models of spatially homogeneous and anisotropic cosmologies. Spatially homogeneous and anisotropic cosmological models play a significant role in the description of large scale behavior of the universe and such models have been widely studied many authors in search of relativistic picture of the early universe. Mohanty \textit{et al.} (2003), Pradhan et al. (2011), Kumar \& Yadav (2011), Yadav  \& Saha (2012) are the some authors who have been extensively investigated an anisotropic Bianchi type-I, Bianchi type-III, Bianchi type-V dark energy models with the usual perfect fluid. Recently, Sarkar \& Mahanta (2013) have studied a homogeneous and anisotropic axially symmetric Bianchi type-I universe filled with matter and holographic dark energy energy components by assuming deceleration parameter to be a constant, along with the correspondence between the holographic dark energy models with the quintessence dark energy and reconstruction of the Quintessence potential and the dynamics of the quintessence scalar field in an accelerated expansion of the universe. Mishra \textit{et al.} (2017) who have investigated the anisotropic behavior of the accelerating universe with the matter field of two non-interacting fluids usual string fluid and dark energy fluid and achieve a physically reasonable clarification using hybrid scale factor, which is generated by a time-varying deceleration parameter and observed that the string fluid dominates the universe at early deceleration phase but does not influence nature of cosmic dynamics significantly at late phase, whereas the dark energy fluid dominates the universe in present time, which is in accordance with the observations results. Chirde \& Shekh (2018a; 2018b) have investigated the dynamics of Bianchi type-I space--time filled with two minimally interacting fields, matter and Holographic dark energy components with volumetric power and exponential expansion laws and the transition between general relativity and quantum gravity using quark and strange quark matter respectively towards the gravitational field equations for the linear form of $f(T)$ gravity and observed that both models are at late times turn out to be flat Universe, the power law model has to begin with singular and stable but at spreading out it is unstable while exponential model is free from any types of singularities and stable throughout the expansion. In the same way the models represents its spreading as accelerated with inflationary era in the early and the very late time of the Universe. For suitably large time resulting model expects that the anisotropy of the Universe will damp out and the Universe will become isotropic. Very recently, Pawar \textit{et al.} (2019) the author who have studied the modified holographic Ricci dark energy model in $f(R,T)$ theory of gravity by using anisotropy Bianchi type Universe and observed that the model has no physical singularity, the Universe is expanding and accelerating exponentially.

\section{Preliminary definitions and equations of motion of $\boldsymbol{f}\boldsymbol{(}\boldsymbol{T}\boldsymbol{)}$ gravity:}\label{1}
In this section we give a brief description of $f(T)$ gravity and a detailed derivation of its field equations.
Let us define the notations of the Latin subscript as these related to the tetrad field and the greek one related to the space-time coordinates. For a general space-time metric, we can define the line element as
\begin{equation}\label{e1} 
dS^{2} =g_{\mu \nu } dx^{\mu } dx^{\nu } ,                   
\end{equation} 
where $g_{\mu \nu } $ are the components of the metric which is symmetric and possesses ten degrees of freedom. One can describe the theory in the space-time or in the tangent space, which allows us to rewrite the line element which can be converted to the Minkowski's description of the transformation called tetrad (which represent the dynamic fields of the theory), as follows

\begin{equation}\label{e2} 
dS^{2} =g_{\mu \nu } dx^{\mu } dx^{\nu } =\eta _{ij} \theta ^{i} \theta ^{j} ,          
\end{equation} 
\begin{equation} \label{e3} 
dx^{\mu } =e_{i}^{\mu } \theta ^{i} ,{\rm \; \; \; }\theta ^{i} =e_{\mu }^{i} dx^{\mu } ,          
\end{equation} 
 where $\eta _{ij} $ is a metric on Minkowski space-time and $\eta _{ij} =diag[1,-1,-1,-1]$ and $e_{i}^{\mu } e_{\nu }^{i} =\delta _{\nu }^{\mu } $ or $e_{i}^{\mu } e_{\mu }^{j} =\delta _{i}^{j} $. The root of metric determinant is given by $\sqrt{-g} =\det [e_{\mu }^{i} ]=e$. For a manifold in which the Riemann tensor part without the torsion terms is null (contribution of the Levi-Civita connection) and only the non-zero torsion terms exist, the Weitzenbocks connection components are defined as 
\begin{equation} \label{e4} 
\Gamma ^{\alpha } {}_{\mu \nu } =e_{i}^{\alpha } \partial _{\nu } e_{\mu }^{i} =-e_{\mu }^{i} \partial _{\nu } e_{i}^{\alpha } .          
\end{equation} 

which has a zero curvature but nonzero torsion. The main geometrical objects of the space-time are constructed from this connection. Through the connection, the components of the tensor torsion are defined by the antisymmetric part of this connection as
\begin{equation} \label{e5} 
T^{\alpha } {}_{\mu \nu } =\Gamma ^{\alpha } {}_{\mu \nu } -\Gamma ^{\alpha } {}_{\nu \mu } =e_{i}^{\alpha } \left(\partial _{\mu } e_{\nu }^{i} -\partial _{\mu } e_{\mu }^{i} \right),         
\end{equation} 
The difference between the Levi-Civita and Weitzenbock connections is a space-time tensor, and is known as the contorsion tensor:
\begin{equation} \label{e6} 
K^{\mu \nu } {}_{\alpha } =\left(-\frac{1}{2} \right)\left(T^{\mu \nu } {}_{\alpha } +T^{\nu \mu } {}_{\alpha } -T^{\mu \nu } {}_{\alpha } \right).         
\end{equation} 
In order to make more clear the definition of the scalar equivalent to the curvature scalar of RG, we first define a new tensor $S^{\mu \nu } {}_{\alpha } $, constructed from the components of the tensors torsion and contorsion tensor as 
\begin{equation} \label{e7} 
S^{\mu \nu } {}_{\alpha } =\left(\frac{1}{2} \right)\left(K^{\mu \nu } {}_{\alpha } +\delta _{\alpha }^{\mu } T^{\beta \nu } {}_{\beta } -\delta _{\alpha }^{\nu } T^{\beta \mu } {}_{\beta } \right).         
\end{equation} 
The torsion scalar $T$ is 
\begin{equation} \label{e8} 
T=T^{\alpha } {}_{\mu \nu } {\rm \; }S^{\mu \nu } {}_{\alpha } .            
\end{equation} 
Now, we define the action by generalizing the TG i.e. $f(T)$ theory as
\begin{equation} \label{e9} 
S=\int \left[T+f(T)+L_{matter} \right] {\rm \; }e{\rm \; }d^{4} x.          
\end{equation} 
Here, $f(T)$ denotes an algebraic function of the torsion scalar \textit{T}. Making the functional variation of the action (9) with respect to the tetrads, we get the following equations of motion
\begin{equation} \label{e10} 
S_{\mu }^{\nu \rho } \partial _{\rho } Tf_{TT} +\left[e^{-1} e_{\mu }^{i} \partial _{\rho } \left(ee_{i}^{\alpha } S_{\alpha }^{\nu \rho } \right)+T^{\alpha } {}_{\lambda \mu } S_{\alpha }^{\nu \lambda } \right]f_{T} +\frac{1}{4} \delta _{\mu }^{\nu } \left(f\right)=4\pi (T_{\mu }^{\nu } {}^{(CS)} +\bar{T}_{\mu }^{\nu } {}^{(DE)} ).    
\end{equation} 
The field equation (10) is written in terms of the tetrad and partial derivatives and appears very different from Einstein's equation.\\
where $T_{\mu }^{\nu } $ is the energy momentum tensor, $f_{T} =df(T)/dT$, $f_{TT} =d^{2} f(T)/dT^{2} $ and by setting $f(T)=a_{0} =$ constant the equations of motion (10) are the same as that of the Teleparallel Gravity with a cosmological constant, and this is dynamically equivalent to the GR. These equations clearly depend on the choice made for the set of tetrads. $T_{\mu }^{\nu } {}^{(CS)} $ and $\bar{T}_{\mu }^{\nu } {}^{(DE)} $ respectively, denote the contribution to energy--momentum tensor from one-dimensional cosmic string and dark energy. \\
In recent years, Cosmologists have taken considerable interest in the study of cosmic strings due to the challenging problem to determine the exact physical scenario at very early stages of the formation of the Universe. Since, they believed that string plays an important role in the description of the Universe in the early stages of evolution (these arise during the phase transition after the big-bang explosion as the temperature decreased below some critical temperature as predicted by grand unified theories) (Kibble  1976). Cosmic string is topologically stable objects that might be found during a phase transition in the early universe and give rise to density perturbations leading to the formation of galaxies (Zeldovich 1980). The innovative work in the formation of the energy-momentum tensor for classical massive strings was done by Letelier (1983), who used this idea to obtain a cosmological solution for Bianchi-I and Kantowski-Sachs space-times by considering the massive strings to be formed by geometric strings with particle attached along its extension.
As in our investigations, one matter content is string cloud fluid. So that the energy momentum tensor for cloud string is defined as 
\begin{equation} \label{e11} 
T_{\mu }^{\nu } {}^{(CS)} =(\rho +p){\rm \; }u^{\nu } u_{\mu } -p{\rm \; }g_{\mu }^{\nu } +\lambda x^{\nu } x_{\mu } ,         
\end{equation} 
where $u^{\nu }$ is the four velocity of the string cloud, $x^{\nu } $ is the normal space-like four-vector, the pressure \textit{p} for this fluid is taken to be isotropic, $\rho $ and $\lambda $ are the proper energy density for a cloud of strings and the tension density of the string cloud respectively. Here we consider the string source is along \textit{z}-axis, which is the axis of symmetry, an orthonormalizsation of $u^{\nu } $ and $x^{\nu } $ is given by 
\begin{equation} \label{e12} 
u^{\nu } u_{\nu } =1, u^{\nu } x_{\nu } =0 , x^{\nu } x_{\nu } =-1,          
\end{equation} 
where $u^{\nu } =\left(0,0,0,1\right)$ and $x^{\nu } =\left(A^{-1} ,0,0,0\right)$. \\
Saha \textit{et al.} (2010) investigated Bianchi type-I cosmological model in the presence of a magnetic flux along with a cosmic string using some tractable assumptions regarding the parameters entering the model and obtained the analytical results which are supplemented with numerical and qualitative analysis describing the evolution of the Universe for different values of the parameters. Sahoo (2015) have studied the non-singular, non-rotating and expanding LRS Bianchi type-I cosmological model in modified gravity with cosmic string and observed that the transition of the universe from decelerated phase to the accelerated phase at late times in accordance with the observations of modern cosmology. The existence of the late time acceleration of the universe with string fluid as source of matter in anisotropic Heckmann-Suchking space-time by using high red shift $(0.3\le z\le 1.4)$ SNe-Ia data of observed complete magnitude along with their possible error from Union 2.1 compilation. It is found that the best fit values for $\left(\Omega _{m} \right)_{0} ,{\rm \; \; }\left(\Omega _{\Lambda } \right)_{0} ,{\rm \; \; }\left(\Omega _{\sigma } \right)_{0} $ and $(q)_{0} $ are 0.2820, 0.7177, 0.0002 \& -0.5793 respectively offered Goswami \textit{et al}. (2016).

The energy--momentum tensor for isotropic dark energy can be described as
\[\bar{T}_{\mu }^{\nu } {}^{(DE)} =diag[-\rho _{\Lambda } ,{\rm \; \; }p_{\Lambda } ,{\rm \; \; }p_{\Lambda } ,{\rm \; \; }p_{\Lambda } {\rm \; }]\] 
\begin{equation} \label{e13} 
=diag[-1,{\rm \; }\omega _{\Lambda } ,{\rm \; }\omega _{\Lambda } ,{\rm \; }\omega _{\Lambda } {\rm \; }]{\rm \; }\rho _{\Lambda } ,        
\end{equation} 
where $\omega _{\Lambda } $ is the equation of state parameter for dark energy and $\rho _{\Lambda } $ is the dark energy density.
\section{ Field equations for Bianchi type-I model:}\label{2}
Let us first establish the equations of motion of a set of diagonal tetrads using the Cartesian coordinate metric, for describing models of Bianchi type-I, as
\begin{equation} \label{e14} 
ds^{2} =dt^{2} -A^{2} dx^{2} -B^{2} dy^{2} -C^{2} dz^{2} ,          
\end{equation} 
Let us choose the following set of diagonal tetrads related to the metric (14)
\begin{equation} \label{e15} 
[e^{\nu } {}_{\mu } ]=diag[1,{\rm \; }A,{\rm \; }B,{\rm \; }C].           
\end{equation} 
The determinant of the matrix (14) is 
\begin{equation} \label{e16} 
e={\rm \; }ABC.            
\end{equation} 
The components of the tensor torsion (5) for the tetrads (15) are given by
\begin{equation} \label{e17} 
T^{1} {}_{01} =\frac{\dot{A}}{A} , T^{2} {}_{02} =\frac{\dot{B}}{B} ,  T^{3} {}_{03} =\frac{\dot{C}}{C} . 
\end{equation} 
and the components of the corresponding tensor contorsion are
\begin{equation} \label{e18} 
K^{01} {}_{1} =\frac{\dot{A}}{A} , K^{02} {}_{2} =\frac{\dot{B}}{B} ,  K^{03} {}_{3} =\frac{\dot{C}}{C} ,  S_{3} {}^{30} =\frac{1}{2} \left(\frac{\dot{A}}{A} +\frac{\dot{B}}{B} \right).     
\end{equation} 
The components of the tensor $S_{\mu } {}^{\mu \nu } $, in (7), are given by
\begin{equation} \label{e19} 
S_{1} {}^{10} =\frac{1}{2} \left(\frac{\dot{B}}{B} +\frac{\dot{C}}{C} \right), S_{2} {}^{20} =\frac{1}{2} \left(\frac{\dot{A}}{A} +\frac{\dot{C}}{C} \right), S_{3} {}^{30} =\frac{1}{2} \left(\frac{\dot{A}}{A} +\frac{\dot{B}}{B} \right).      
\end{equation} 
The corresponding torsion scalar (8) is given by
\begin{equation} \label{e20} 
T=-2{\rm \; }\left(\frac{\dot{A}}{A} \frac{\dot{B}}{B} +\frac{\dot{A}}{A} \frac{\dot{C}}{\dot{C}} +\frac{\dot{B}}{B} \frac{\dot{C}}{C} \right).          
\end{equation} 
Now, the field equations in the framework of $f(T)$gravity (10) for a two-fluid cosmic string (11) and dark energy (13) Bianchi type-I space-time (14) are obtained as
\begin{equation} \label{e21} 
f+4f_{T} \left\{\frac{\dot{A}}{A} \frac{\dot{B}}{B} +\frac{\dot{A}}{A} \frac{\dot{C}}{C} +\frac{\dot{B}}{B} \frac{\dot{C}}{C} -\frac{\alpha ^{2} }{2A^{2} } \right\}=\left(\rho +\rho _{\Lambda } \right),        
\end{equation} 
\begin{equation} \label{e22} 
f+2f_{T} \left\{\frac{\ddot{B}}{B} +\frac{\ddot{C}}{C} +\frac{\dot{A}}{A} \frac{\dot{B}}{B} +\frac{\dot{A}}{A} \frac{\dot{C}}{C} +2\frac{\dot{B}}{B} \frac{\dot{C}}{C} \right\}+2\left(\frac{\dot{B}}{B} +\frac{\dot{C}}{C} \right)\dot{T}f_{TT} =-\left(p+\omega _{\Lambda } \rho _{\Lambda } \right),     
\end{equation} 
\begin{equation} \label{e23} 
f+2f_{T} \left\{\frac{\ddot{A}}{A} +\frac{\ddot{C}}{C} +\frac{\dot{A}}{A} \frac{\dot{B}}{B} +2\frac{\dot{A}}{A} \frac{\dot{C}}{C} +\frac{\dot{B}}{B} \frac{\dot{C}}{C} \right\}+2\left(\frac{\dot{A}}{A} +\frac{\dot{C}}{C} \right)\dot{T}f_{TT} =-\left(p+\omega _{\Lambda } \rho _{\Lambda } \right),     
\end{equation} 
\begin{equation} \label{e24} 
f+2f_{T} \left\{\frac{\ddot{A}}{A} +\frac{\ddot{B}}{B} +2\frac{\dot{A}}{A} \frac{\dot{B}}{B} +\frac{\dot{A}}{A} \frac{\dot{C}}{C} +\frac{\dot{B}}{B} \frac{\dot{C}}{C} -\frac{\alpha ^{2} }{A^{2} } \right\}+2\left(\frac{\dot{A}}{A} +\frac{\dot{B}}{B} \right)\dot{T}f_{TT} =-\left(p+\lambda +\omega _{\Lambda } \rho _{\Lambda } \right),    
\end{equation} 
where the dot ($\cdot $) denotes the derivative with respect to cosmic time \textit{t}.
In the particular case where $f(T)=T-2\Lambda $, the equations (21) - (24) are identical to that of the general theory of relativity. 
Finally, here we have four differential equations with eight unknowns namely $A, {\rm \; }B, {\rm \; }f, {\rm \; }p, {\rm \; }\rho ,\lambda, \rho _{\Lambda }, \omega _{\Lambda } $. The solution of these equations is discussed in next section. In the following we define some kinematical quantities of the space-time.
We define average scale factor and volume respectively as
\begin{equation} \label{e25} 
R^{3} =V=ABC.           
\end{equation} 
Another important dimensionless kinematical quantity is the mean deceleration parameter which tells whether the Universe exhibits accelerating volumetric expansion or not is
\begin{equation} \label{e26} 
q=-1+\frac{d}{dt} \left(\frac{1}{H} \right),           
\end{equation} 
for $-1\le q<0$, $q>0$ and $q=0$ the Universe exhibit accelerating volumetric expansion, decelerating volumetric expansion and expansion with constant-rate respectively.
The mean Hubble parameter, which expresses the volumetric expansion rate of the Universe, given as
\begin{equation} \label{e27} 
H=\frac{1}{3} \left(H_{1} +H_{2} +H_{3} \right),           
\end{equation} 
where $H_{1} $, $H_{2} $ and ${\rm \; }H_{3} $ are the directional Hubble parameter in the direction of \textit{x, y} and \textit{z}-axis respectively.
Using equations (25) and (27), we obtain
\begin{equation} \label{e28} 
H=\frac{1}{3} \frac{\dot{V}}{V} =\frac{1}{3} \left(H_{1} +H_{2} +H_{3} \right)=\frac{\dot{R}}{R} .         
\end{equation} 
To discuss whether the Universe either approach isotropy or not, we define an anisotropy parameter as
\begin{equation} \label{e29} 
A_{m} =\frac{1}{3} \sum _{i=1}^{3}\left(\frac{H_{i} -H}{H} \right) ^{2} .           
\end{equation} 
The expansion scalar and shear scalar are defined as follows
\begin{equation} \label{e30} 
\theta =u_{;\mu }^{\mu } =\frac{\dot{A}}{A} +\frac{\dot{B}}{B} +\frac{\dot{C}}{C} ,           
\end{equation} 
\begin{equation} \label{e31} 
\sigma ^{2} =\frac{3}{2} H^{2} A_{m} .            
\end{equation} 
\section{Solution of the field equations:}\label{3}
In order to solve the field equations completely, we first assume that the matter and holographic dark energy components i.e. the energy momentum tensors of the two sources interact minimally and conserved separately.
The energy conservation equation of the matter leads to\textbf{}
\begin{equation} \label{e32} 
\left(\dot{\rho }\right)+\frac{\dot{V}}{V} \left(\rho \right)=0.           
\end{equation} 
\begin{equation} \label{e33} 
\rho =\frac{1}{V} .            
\end{equation} 
In this section we find exact solutions of field equations using some physical quantities for linear $f(T)$ gravity i.e.
\begin{equation} \label{e34} 
f(T)=T.            
\end{equation} 
Now subtracting (22) from (23), we get
\begin{equation} \label{e35} 
\frac{d}{dt} \left(\frac{\dot{A}}{A} -\frac{\dot{B}}{B} \right)+\left(\frac{\dot{A}}{A} -\frac{\dot{B}}{B} \right)\frac{\dot{V}}{V} =0,          
\end{equation} 
which on integration gives
\begin{equation} \label{e36} 
\frac{A}{B} =k_{2} \exp \left[k_{1} \int \frac{dt}{V}  \right],           
\end{equation} 
where $k_{1} $and$k_{2} $are constants of integration.
In view of equation (25), we write \textit{A }and \textit{B} in the explicit form 
\begin{equation} \label{e37} 
A=D_{1} V^{\frac{1}{3} } \exp \left(\chi _{1} \int \frac{1}{V} dt \right),           
\end{equation} 
\begin{equation} \label{e38} 
B=D_{2} V^{\frac{1}{3} } \exp \left(\chi _{2} \int \frac{1}{V} dt \right),          
\end{equation} 
where $D_{i} (i=1,\; 2)$ and $\chi _{i} (i=1,\; 2)$ satisfy the relation $D_{1} D_{2}^{2} =1$ and $\chi _{1} +2\chi _{2} =0$. \\
\\
\textbf{Exact matter dominated power law solution of the field equation}
\\
\\
Since field equations are highly nonlinear, an extra condition is needed to solve the system completely. So, we have chosen the scale factor of the form
\begin{equation} \label{e39} 
V=R^{3} =t^{b}  
\end{equation} 
where \textit{b} be the any positive constant. 
The motivation to choose such scale factor is that the universe has accelerated expansion at present and decelerated expansion in the past. Here we are considering the minimally interacting but conserve separately matter and holographic dark energy components.
Using equations (37), (38) and (39), we obtain the metric potentials as follows:
\begin{equation} \label{e40} 
A=D_{1} t^{\frac{b}{3} } \exp \left\{\frac{\chi _{1} }{(1-b)} t^{1-b} \right\},          
\end{equation} 
\begin{equation} \label{e41} 
B=D_{2} t^{\frac{b}{3} } \exp \left\{\frac{\chi _{2} }{(1-b)} t^{1-b} \right\},          
\end{equation} 
and 
\begin{equation} \label{e42} 
C=\frac{1}{D_{1} D_{2} } t^{\frac{b}{3} } \exp \left\{\frac{\chi _{2} }{(1-b)} t^{1-b} \right\}.          
\end{equation} 
From the equations (40) to (42), it is observed that the metric potentials \textit{A, B }and \textit{C} are the product of power and exponential form and increase indefinitely with the passage of time.
With the help of equations (40) to (42), our spatially homogeneous and anisotropic Bianchi type-I space-time filled with cosmic string and holographic dark energy fluid within the framework of $f(T)$ gravity becomes 
\begin{equation} \label{e43} 
ds^{2} =dt^{2} -D_{1}^{2} t^{\frac{2b}{3} } \exp \left\{\frac{2\chi _{1} }{(1-b)} t^{1-b} \right\}dx^{2} -D_{2}^{2} t^{\frac{2b}{3} } \exp \left\{\frac{2\chi _{2} }{(1-b)} t^{1-b} \right\}dy^{2} -\frac{1}{D_{1}^{2} D_{2}^{2} } t^{\frac{2b}{3} } \exp \left\{\frac{2\chi _{2} }{(1-b)} t^{1-b} \right\}dz^{2} .              
\end{equation} 
All the metric potentials in the derived model are vanishes at $t=0$. Hence, the model has no initial singularity. Afterwards, increase indefinitely with the passage of time, which is in complete agreement with the Big-Bang model of the Universe and the model is similar to the investigation those of Chirde \& Shekh (2018).

\section{Kinematical parameters:}

Spatial volume and Average scale factor,
\begin{equation} \label{e44} 
V=R^{3} =t^{b}  
\end{equation} 
The expansion scalar,
\begin{equation} \label{e45} 
\theta =\frac{b}{t} ,             
\end{equation} 
The mean Hubble parameter, 
\begin{equation} \label{e46} 
H=\frac{b}{3t} .            
\end{equation} 
We observe that the spatial volume and average scale factor both are die out at $t\to 0$. Thus, no singularity exists at $t\to 0$ in the resultant model. The model starts evolving with a zero volume at $t\to 0$. The behavior is shown in FIG-1. The expansion scalar decreases as time increases and the mean Hubble parameter is initially large at $t\to 0$, and null at $t\to \infty $. The expansion scalar $\theta \to 0$ as $t\to \infty $ which indicates that the universe is expanding with increase of time and the rate of expansion decreases with the increase of time. This suggests that at initial stage of the universe, the expansion of the model is much more faster and then slows down for later time i.e. the evolution of the universe starts with infinite rate, and with expansion, it declines (see FIG-2).\\

\begin{figure}
  \centerline{\includegraphics[scale=0.30]{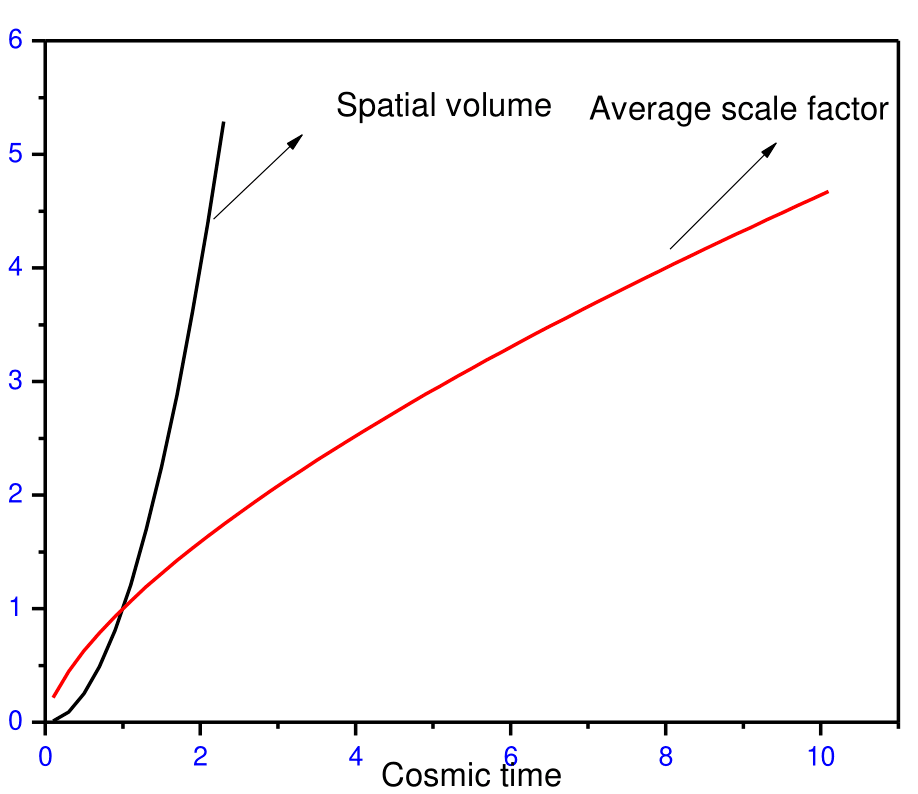}}   
\caption{Graphical representation of average scale factor and spatial volume versus cosmic time with appropriate choice of constant $b=2$.}
\end{figure}
\begin{figure}
  \centerline{\includegraphics[scale=0.30]{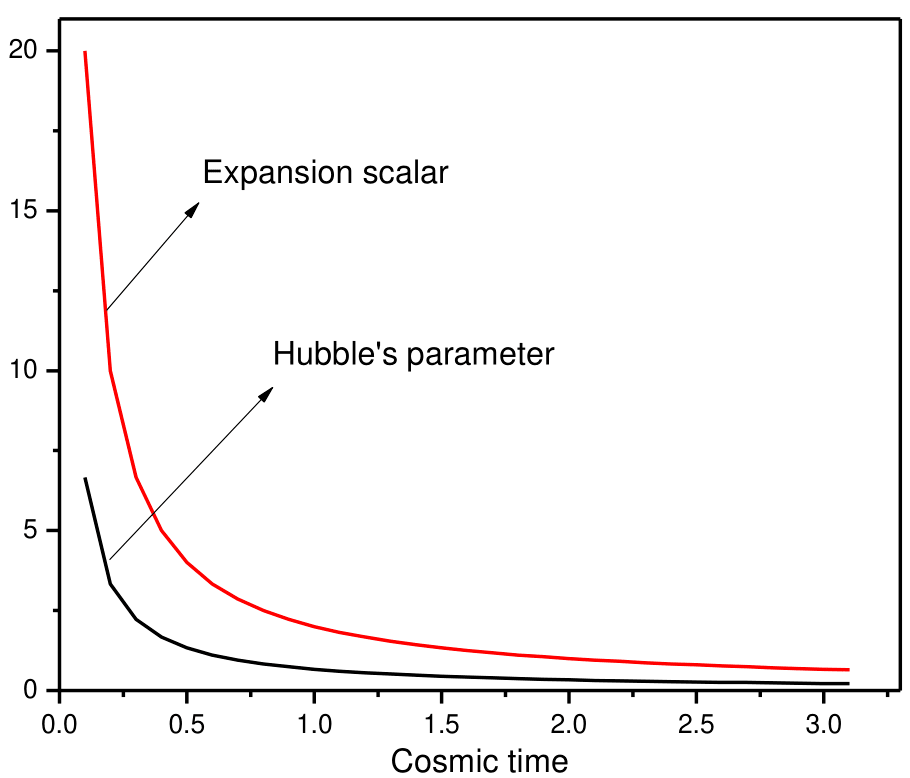}}   
\caption{Graphical representation of Hubble's parameter and expansion scalar versus cosmic time with appropriate choice of constant  $b=2$.}
\end{figure}
An anisotropy parameter,
\begin{equation} \label{e47} 
A_{m} =\frac{18\chi _{2} {}^{2} }{b^{2} t^{2(b-1)} } .           
\end{equation} 
The shear scalar,
\begin{equation} \label{e48} 
\sigma ^{2} =\frac{9\chi _{2} {}^{2} }{3t^{2b} } .            
\end{equation} 
The anisotropic parameter and the shear scalar both are the inverse function of time. Thus the nature of the anisotropic parameter is varying with the evaluation of the universe. Initially both are in high elevation and at an infinite expansion it is seen that the model has shear free and that of anisotropy parameter disappears in the model. The behavior of the anisotropy parameter and shear scalar versus cosmic time \textit{t }is shown in FIG-3.
\begin{figure}
  \centerline{\includegraphics[scale=0.30]{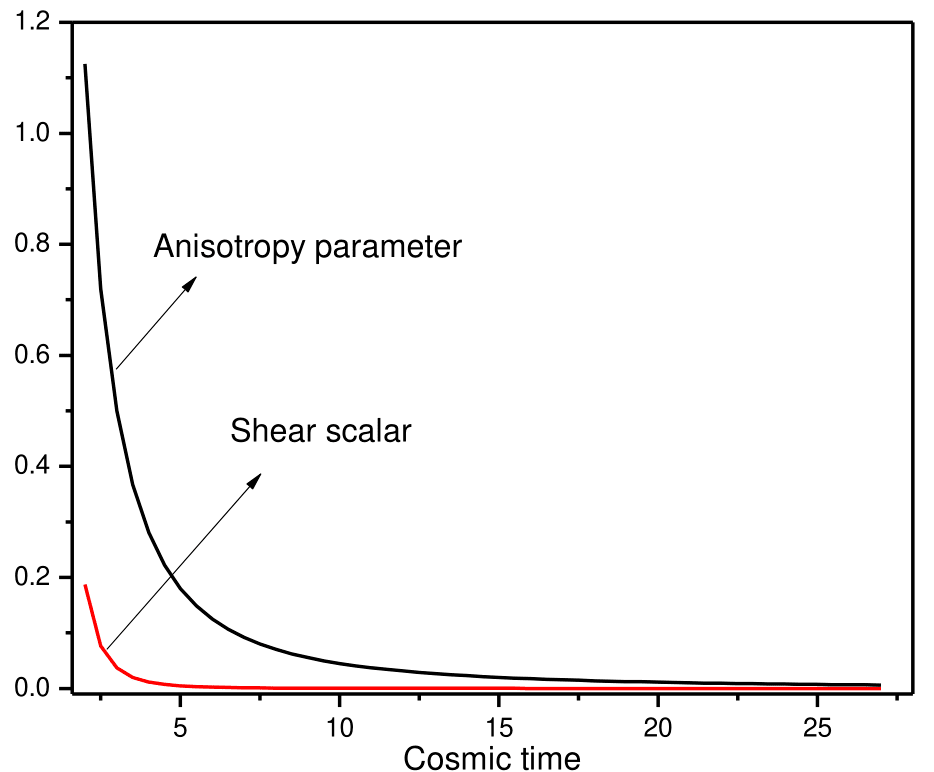}}   
\caption{Graphical representation of anisotropy parameter and shear scalar versus cosmic time with appropriate choice of constants $b=2$ and $\chi _{2} =1$.}
\end{figure}
\section{Physical parameters:}\label{2}

The Torsion scalar for the model becomes 
\begin{equation} \label{e49} 
T=\frac{-6\chi _{1} \chi _{2} }{t^{2b} } -\frac{16b\left(\chi _{1} +\chi _{2} \right)}{3t^{b+1} } -\frac{8b^{2} }{9t^{2} }  
\end{equation} 

\begin{figure}
  \centerline{\includegraphics[scale=0.30]{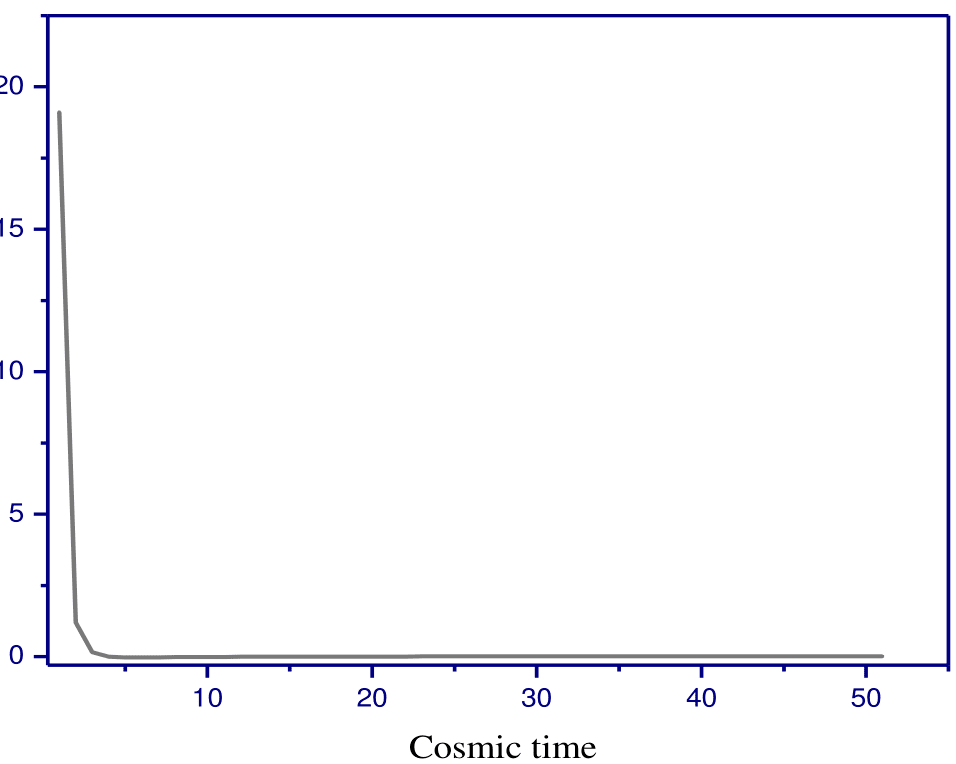}}   
\caption{Graphical representation of Torsion scalar versus cosmic time with appropriate choice of constants $b=2$, $\chi _{1} =-2$ and $\chi _{2} =1$.}
\end{figure}

The Torsion of the Universe is time dependent and decreases very rapidly for the interval of cosmic time $1\le t\le 15.5$ and for $t>15.5$ it attend a very small positive value which is nearly equal to zero, the behavior is clearly shown in FIG-4.
The proper energy density for a cloud of string is
\begin{equation} \label{e50} 
\rho =\frac{1}{t^{b} } .            
\end{equation} 
The energy density of Holographic dark energy is,
\begin{equation} \label{e51} 
\rho _{\Lambda } =\frac{\chi _{2} (\chi _{1} +\chi _{2} )}{t^{2b} } +\left(\frac{b\left(\chi _{1} +3\chi _{2} \right)}{t^{b+1} } \right)-\frac{1}{t^{b} } +\frac{2b^{2} }{9t^{2} } .       
\end{equation} 
\begin{figure}
  \centerline{\includegraphics[scale=0.30]{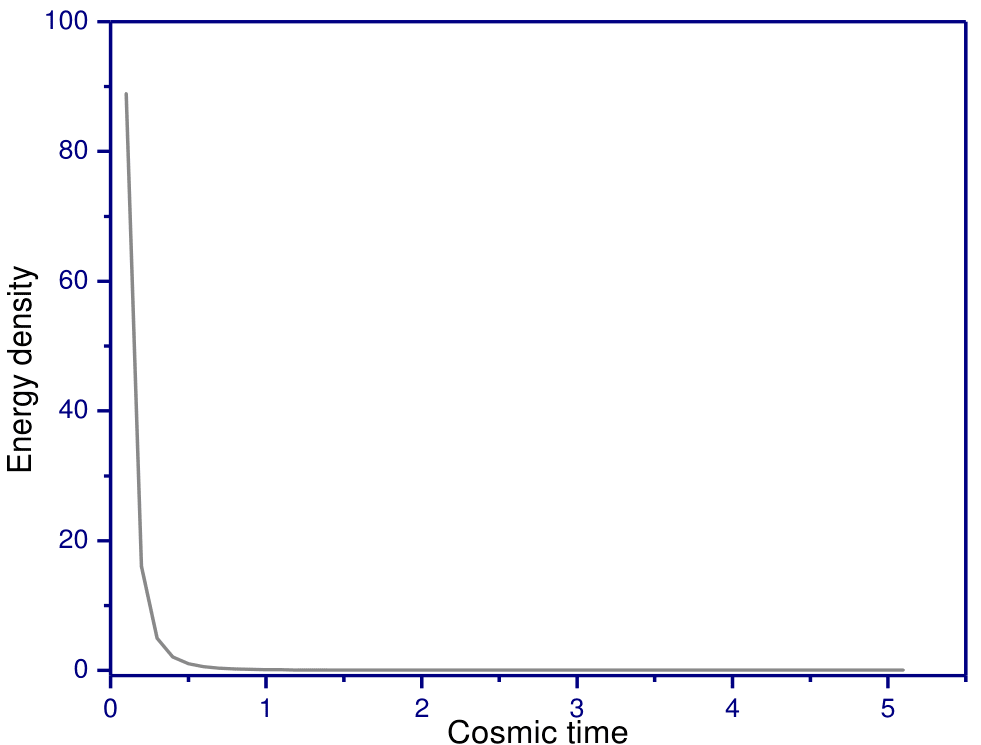}}   
\caption{Graphical representation of energy density versus cosmic time with appropriate choice of constants $b=2$, $\chi _{1} =-2$ and $\chi _{2} =1$.}
\end{figure}
In power law expansion of the Universe, it is observed that the energy density is always positive and decreasing function of cosmic time \textit{t}. At the initial stage the Universe has infinitely large energy density but with the expansion of the Universe it declines and at very large expansion, it is null. The behavior is clearly shown in FIG-5. \\
The equation of state parameter for Holographic dark energy is,
\begin{equation} \label{e52} 
\omega _{\Lambda } =\frac{\left(\frac{\chi _{2} \left(2\chi _{1} -6\chi _{2} -\omega \right)}{t^{2b} } +\frac{b\left(2\chi _{1} -6\chi _{2} -\omega \right)}{3t^{b+1} } \right)}{\left(\frac{\chi _{2} (\chi _{1} +\chi _{2} )}{t^{2b} } +\left(\frac{b\left(\chi _{1} +3\chi _{2} \right)}{t^{b+1} } \right)-\frac{1}{t^{b} } +\frac{2b^{2} }{9t^{2} } \right)} .       
\end{equation} 
Recently a large class of scalar field dark energy models has been given including Quintessence $(\omega _{\Lambda } >-1)$, Phantom $(\omega _{\Lambda } <-1)$ and Quinton (which can cross from the Phantom region to the quintessence region). The Quinton scenario of dark energy is designed to understand the nature of dark energy with $\omega _{\Lambda } $ cross $-1$. Setare \& Saridakis (2015) have studied the dark energy models with the equation of state parameter across ($-1$), which gives a concrete justification for the Quinton paradigm. Some other limits of equation of state parameter are obtained from observational results that came from SNe-Ia data and a combination of SNe-Ia data with CMB anisotropy and Galaxy clustering statistics are $-1.66<\omega _{\Lambda } <-0.62$ and $-1.33<\omega _{\Lambda } <-0.79$ respectively. The latest result in 2009, obtained after a combination of cosmological data sets coming from CMB anisotropy, luminosity distances of high red-shift SNe-Ia, and galaxy clustering constrain the dark energy equation of state parameter to $-1.44<\omega _{\Lambda } <-0.92$. In the derived model, the equation of state parameter is evolving with a positive sign i.e. $\omega _{\Lambda } >0$ where the model behaves as like matter dominated once at early stages for the interval $0.01\le t\le 0.05$, for the interval $0.06\le t\le 1.73$ the model shows quintessence region while at late times at $t>1.73$ it be present in matter dominated once, which may be established from the current accelerated expansion of the universe. From figure-\textit{vi}, we observed that at the initial time there is a quintessence region, and at late time it approaches the cosmological constant scenario. This is a situation in the early universe where the quintessence-dominated universe (Caldwell 2002) may be playing an important role for the equation of state parameter.
If the present model is compared with the experimental results mentioned above, one can conclude that the limit of equation of state parameter provided by Eq. (31) for some interval of time it may be accumulated with the acceptable range of equation of state parameter (see Fig-6). This model confirms the high red-shift supernova experiment.\\

\begin{figure}
  \centerline{\includegraphics[scale=0.30]{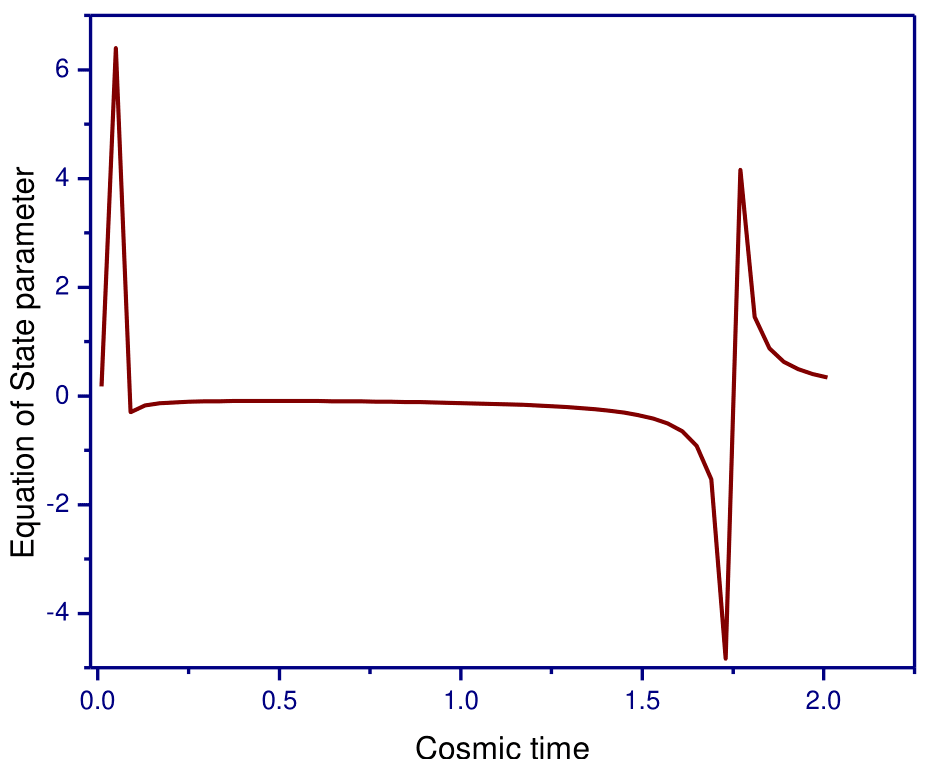}}   
\caption{Graphical representation of Equation of State parameter versus cosmic time with appropriate choice of constants $b=2$, $\chi _{1} =-2$, $\chi _{2} =1$ and $\omega =1/3$.}
\end{figure}

The tension density of the string cloud is
\begin{equation} \label{e53} 
\lambda =\left[\frac{\chi _{2} }{t^{b} } +\frac{b}{3t} \right]\left\{\left. \frac{\omega +4\chi _{2} -2\chi _{1} }{t^{b} } \right\}-\frac{\omega }{t^{b} } -2\frac{\chi _{1} }{t^{b} } \left[\frac{\chi _{1} }{t^{b} } +\frac{b}{3t} \right]\right. .      
\end{equation} 
As the cosmologists have taken considerable interest in the study of cosmic strings. Since, they believed that string plays an important role in the description of the Universe in the early stages of evolution i.e. arise during the phase transition after the big bang explosion as the temperature decreased below some critical temperature as predicted by grand unified theories also which is a topologically stable objects that might be found during a phase transition in the early universe, Letelier (1983) pointed out that tension density of the string cloud may be positive or negative. In our investigation it is initially positive hence the strings exist in the early universe occupying the small universe, which is in agreement with constraints of CMBR data but with the expansion of the Universe it gradually decreases and remain present with negative value i.e. $\lambda<$ 0 hence, with the expansion the string phase of the Universe disappears, i.e. we have an anisotropic fluid of particles, this behavior is clearly shown in FIG-7.

\begin{figure}
  \centerline{\includegraphics[scale=0.30]{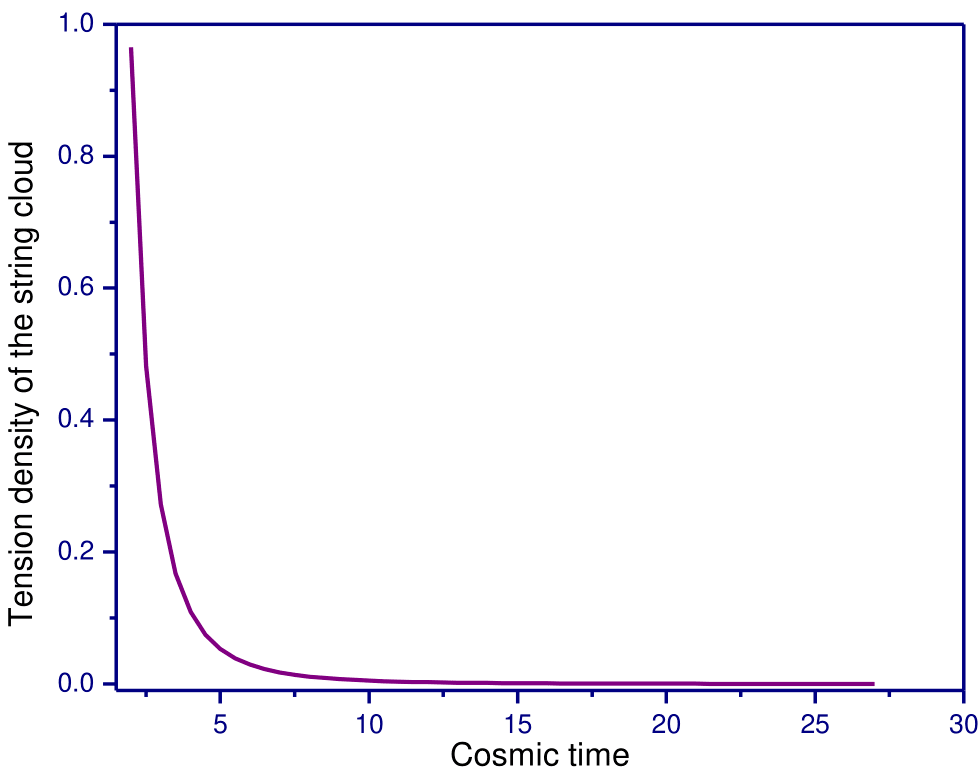}}   
\caption{Graphical representation of tension density of the string cloud versus cosmic time with appropriate choice of constants $b=2$, $\chi _{1} =-2$, $\chi _{2} =1$ and $\omega =1/3$.}
\end{figure}

Stability factor of the Universe,

For the stability of corresponding solutions of the derived Universe, we should check that our Universe is physically acceptable. For this, firstly it is required that the velocity of sound should be less than velocity of light i.e. within the range $0<\upsilon _{s} =\frac{\partial p}{\partial \rho } $.

In our derived model, the sound speeds is obtained as
\begin{equation} \label{e54} 
\upsilon _{s} =\frac{\left(\frac{2b\chi _{2} (2\chi _{2} +\omega )}{t^{2b+1} } +\frac{b(b+1)(2\chi _{2} +\omega )}{3t^{b+2} } \right)}{\left(\frac{2b\chi _{2}^{2} }{t^{2b+1} } +\frac{b}{t^{b+1} } -\frac{b\chi _{2} (b+1)}{t^{b} } -\frac{2b^{2} }{9t^{3} } \right)} .        
\end{equation} 
\begin{figure}
  \centerline{\includegraphics[scale=0.30]{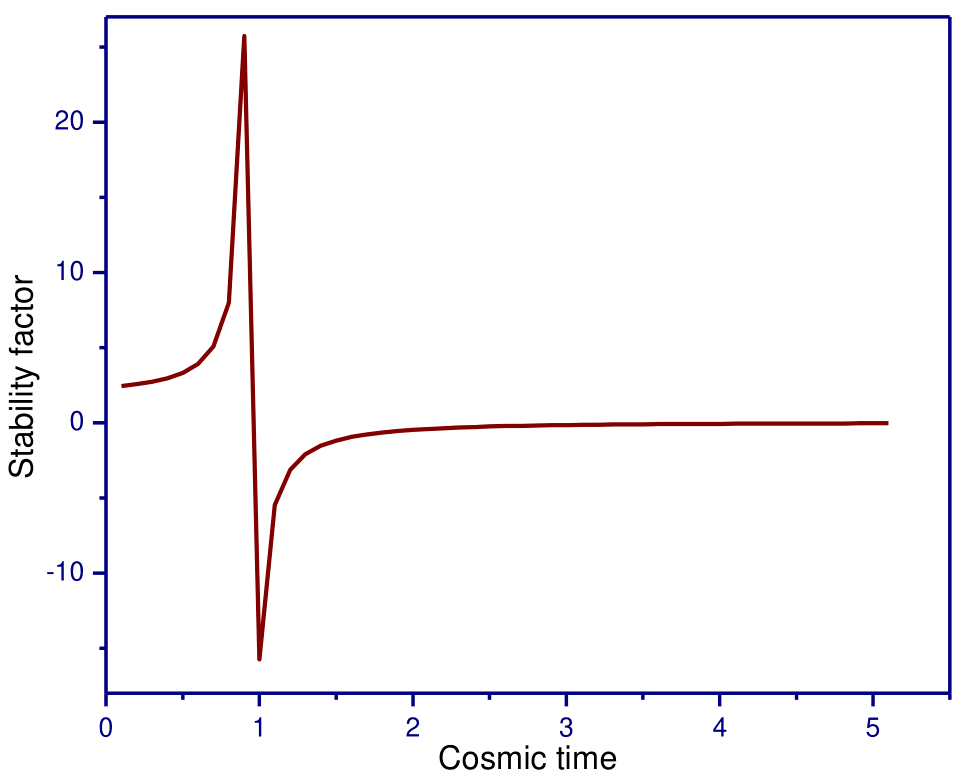}}   
\caption{Graphical representation of stability factor of the Universe versus cosmic time with ape choice of constants $b=2$, $\chi _{1} =-2$, $\chi _{2} =1$ and $\omega =1/3$.}
\end{figure}
From the FIG-8, it is observed that initially for $0<t\le 0.90$ stability factor $\upsilon _{s} >0$ and with the expansion $\upsilon _{s} <0$. Hence the Universe initially stable but with expansion it is unstable. Therefore, on the basis of above discussions and analysis, our corresponding solutions are physically acceptable at initially for small interval of cosmic time.\textbf{}

\section{Conclusion}

The Torsion of the Universe is time varying and falls very rapidly for $1\le t\le 15.5$ and for $t>15.5$ it attends a very small positive value which is nearly equal to zero. The energy density is always positive hence the model is realistic, at an initial stage the Universe has infinitely large energy density but with the expansion of the Universe it declines and at very large expansion, it is null. 

The Equation of State parameter is evolving with a positive sign for this case, the model behaves as like matter dominated once at early stages at an interval $0.01\le t\le 0.05$, for the interval $0.06\le t\le 1.73$ the model shows quintessence region which may be established from the current accelerated expansion of the universe while at late times at $t>1.73$ it is present in matter dominated once. At the initial time there is a quintessence region, and at late time it is baryonic matter. This is a situation in the early universe where the quintessence-dominated universe may be playing an important role for the equation of state parameter. If the present model is compared with the experimental results, it may be accumulated with the acceptable range of equation of state parameter. This model confirms the high red-shift supernova experiment.

Initially the cosmic string has positive, hence the strings exist in the early universe occupying the small universe, which is in agreement with constraints of CMBR data but with the expansion of the Universe it gradually decreases and remain present with negative value i.e. $\lambda<$ 0 hence, with the expansion the string phase of the Universe disappears, i.e. we have an anisotropic fluid of particles.

\end{document}